\documentclass[
    ,final            % use final for the camera ready runs
  ]
  {aipproc}

\layoutstyle{8x11double}
\def\lsim{{\buildrel <\over \sim}}

\begin{document}

\title[]{Large Thermoelectric Effects and Inelastic Scattering in Unconventional
Superconductors}


\author{Mikael Fogelstr\"om}
{address={Applied Quantum Physics, MC2, Chalmers University of Technology, S-412 96 G\"oteborg, Sweden}}
\author{Tomas L\"ofwander}
{address={Institut f\"ur Theoretische Festk\"orperphysik,
Universit\"at Karlsruhe, 76128 Karlsruhe, Germany}}

\begin{abstract}
The thermoelectric coefficient $\eta(T)$ in unconventional
superconductors is enhanced below $T_c$ by intermediate strength
impurity scattering that is intrinsically particle-hole asymmetric.
%The presence of a finite $\eta(T)$
%may be measured as a temperature-gradient induced large magnetic flux
%in a SQUID setup.
We compute $\eta(T)$ for a strong-coupling d-wave superconductor and
investigate the effects of inelastic scattering originating from
electron-boson interactions. We show that $\eta(T)$ is severely
suppressed at temperatures just below $T_c$ by a particle-hole
symmetric inelastic scattering rate.  At lower temperatures inelastic
scattering is frozen out and $\eta(T)$ recovers and regains its large
amplitude. In the limit $T\rightarrow 0$, we have $\eta(T)\sim
\eta_{0} T+{\cal O}[T^3]$, where the slope $\eta_{0}$ contains
information about the Drude plasma frequency, the details of impurity
scattering, and the change in effective mass by electron-boson
interactions. In this limit $\eta(T)$ can be used as a probe,
complementary to the universal heat and charge conductivities, in
investigations of the nature of nodal quasiparticles.
\end{abstract}

\maketitle

%%%%%%%%%%%%%%%%%%%%%%%%%%%%%%%%%%%%%%%%%%%%
%% MAINMATTER
%%%%%%%%%%%%%%%%%%%%%%%%%%%%%%%%%%%%%%%%%%%%
Low-temperature transport measurements have provided a wealth of
information about nodal quasiparticles in unconventional
superconductors \cite{Experiment,Hill04}. Thermal conductivity is of
particular importance, because theory predicts universality in the
sense that the low-temperature asymptotic does not depend on the
properties of the impurity potential \cite{Universal}. This prediction
has been confirmed experimentally \cite{Experiment}. However, there
are some difficulties in analysing the low-temperature thermal
conductivity. First, the leading $T^2$-dependence of the electronic
contribution to $\kappa(T)/T$ is masked by a phonon contribution with
the same $T^2$-dependence \cite{Hill04}. Second, experiments are often
done on ultra-clean samples, which means that the $T^2$ power law of
$\kappa(T)/T$ holds only in a small temperature bracket \cite{LF05}.

In a recent paper we discussed how isotropic elastic scattering by
impurities of intermediate strength, {\it i.e} described by a phase
shift $0<\delta_0 < \frac{\pi}{2}$, gives rise to an electron-hole
asymmetric scattering time and consequently a large non-universal
thermoelectric response \cite{LF04}. A careful study of the
thermoelectric coefficient in the low-temperature regime would reveal
information about the bare elastic scattering rate and potential
strength. This information is hard to extract from thermal
conductivity data. In this report we examine the interplay between
elastic and inelastic scattering and show how it affects the
thermoelectric coefficient.

%%%%%%%%%%%%%%%%%%%%%%%%%%%%%%%%%%%%%%%%%%%%%%%%
\begin{figure}[t]
  \includegraphics[width=0.92\textwidth]{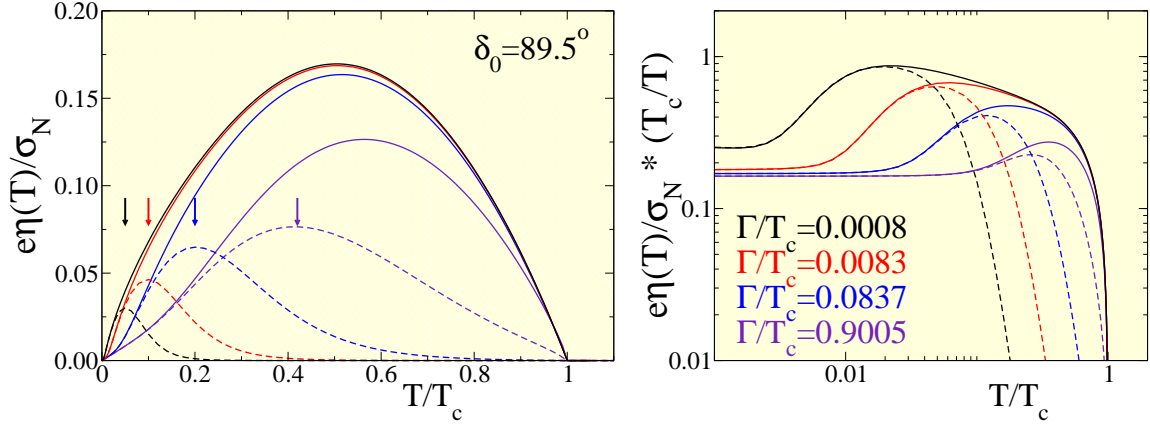}%pdf}
  \caption{Thermoelectric coefficient, $\eta(T)$, calculated for four
orders of magnitude in scattering rate, as indicated in the right
panel. The dashed lines are results of a self-consistent calculation
with the inelastic scattering model introduced in
Ref.~\cite{LF05}. The full lines are the corresponding results
obtained with an {\em effective} model where the inelastic part is
included only as an effective mass via $m^*/m=1+\lambda_{\sf in}$
\cite{LF05}. The temperature dependence obtained with the effective
model is the same as within weak-coupling theory. In the left panel
$\eta(T)$ is scaled by the normal state Drude conductance,
$\sigma_N=e^2 v_f^2 {\cal N}_f/2 \Gamma$, where $v_f, {\cal N}_f,$ and
$\Gamma$ are effective values of respective bare constants. The arrows
indicate the cross-over, $T^*$, between the two scattering modes. In the
right panel we show the temperature dependence of $\eta(T)/T$ with
emphasis on the low-temperature regime where the leading
$T-$dependence is $\eta(T)/T\approx \eta_0$. The constant, $\eta_0$,
is non-universal since it explicitly depends on the ratio of
$\partial\Im \Sigma^R_{0,imp}(\epsilon)/\partial \epsilon|_{\epsilon=0}$
and $\Im \Sigma^R_{3,imp}(\epsilon=0)$.}
\label{Fig1}
\end{figure}
%%%%%%%%%%%%%%%%%%%%%%%%%%%%%%%%%%%%%%%%%%%%%%%%

The thermoelectric coefficient, $\eta(T)$, is defined as
\begin{equation} \delta \vec j_e=-\eta(T) \nabla T =2 {\cal N}_f
\int d{\vec p_f} \int \frac{d \epsilon}{4\pi i} \,e\, {\vec v_f}
\delta g^K.
\end{equation} 
The quasiclassical propagator $\delta{\hat g^K}$ has a closed form in
which the self-consistently computed equilibrium Green function ${\hat
g^R_0}$ and the self-energy $\hat\Sigma^R$ serve as input (see Graf
{\it et al.} \cite{Universal} and Ref. \cite{LF04} for details).  In
the present study we consider a composite self-energy in particle-hole
space $\hat\Sigma^R=\hat\Sigma^R_{\rm imp}+\hat\Sigma^R_{\rm in}$,
where the impurity self-energy $\hat\Sigma^R_{\rm imp}$ is diagonal
with $\Sigma^R_{3,\rm imp}$ and $\Sigma^R_{0,\rm imp}$ being its
particle-hole symmetric and anti-symmetric parts, respectively. The
self-energy $\hat \Sigma^R_{\rm in}$ includes the effects of inelastic
electron-boson scattering, but we also assume that this interaction
mediates the pairing. Below, $\hat W^R$ is the off-diagonal component,
{\it i.e} the usual strong-coupling function related to the energy
dependent gap as $\hat \Delta^R(\epsilon)=\hat
W^R(\epsilon)/Z(\epsilon)$.  Contrary to the impurity self-energy the
diagonal part of the inelastic self-energy is
particle-hole symmetric as $\Sigma^R_{0,\rm in}=0$. Finally, $Z(\epsilon)$ is the
energy-renormalization function defined by the scattering renormalized
energy $\tilde
\epsilon^R=Z^R(\epsilon)\,\epsilon=\epsilon-\Sigma^R_{3,\rm
imp}-\Sigma^R_{3,\rm in}$.  This model is the same as we used in our
study of the thermal conductivity \cite{LF05}. With this input we use
Ref.~\cite{LF04} and write down the response function
\begin{eqnarray}
\eta_{ij}(T)= -\frac{e}{4 T^2}
\int d\epsilon\,\,\epsilon\,{\rm sech}^2\frac{\epsilon}{2 T}
\int d{\vec p_f} [v_{f,i} v_{f,j}] &&
\nonumber
\\
\times\frac{{{\cal N}}({\vec p_f},\epsilon)\,\Im\,\Sigma^R_{0,imp}(\epsilon)}
{\left[\Re\,\Omega^R({\vec p_f};\epsilon)\right]^2-\left[\Im\,\Sigma^R_{0,imp}(\epsilon)\right]^2},&&
\label{eq:L12}\end{eqnarray}
where ${\cal N}({\vec p_f},\epsilon)= - {\cal N}_f \Im\,\left[\tilde
\epsilon^R/\Omega^R({\vec p_f};\epsilon)\right]$ is the density of
states, and $\Omega^R=\sqrt{|W^R({\vec p_f},\epsilon)|^2 -(\tilde
\epsilon^R)^2}$. Note that $\eta(T)$ is directly proportional to the
imaginary part of the particle-hole asymmetric part
$\Im\,\Sigma^R_{0,imp}$, which is an odd function of energy.

In Figure~\ref{Fig1} we show $\eta(T)$ for a strong-coupling d-wave
superconductor. The results are compared with an effective
model-calculation where only the mass renormalization by inelastic
scattering is accounted for, as described in Ref.~\cite{LF05}.  As
seen in Figure~\ref{Fig1}, inelastic scattering affects $\eta(T)$ much
in the same way as it does the thermal conductivity reported in
Ref.~\cite{LF05}, {\it i.e} it dominates at temperatures $T \leq T_c$
but freezes out at low $T\ll T_c$ where instead elastic scattering
limits the transport. We separate the two regimes where the two types
of scattering dominate by introducing a cross-over temperature $T^*$,
defined as the temperature where $\eta(T)$ has its maximium. The value
of $T^*$ depends on the elastic scattering rate $\Gamma$ for a given
electron-boson coupling spectra. The suppression of $\eta(T)$ in the
interval $T^*\lsim T\leq T_c$ comes from that in our model inelastic
scattering does not break particle-symmetry and hence reduces the
weight of the kernel in the integral in Eq.~(\ref{eq:L12}).

In the low-$T$ regime $T\ll |\Im \Sigma^R_{3,imp}(0)|$, we obtain
\begin{equation}
\frac{\eta(T)}{T} =
e\,\frac{\pi^2}{3} \frac{2 {\cal N}_f v_f^2}{\pi \mu\Delta_0}
\left|\frac{\partial_{\epsilon} \Im \Sigma^R_{0,imp}(\epsilon)}{\Im \Sigma^R_{3,imp}(\epsilon)}\right|_{\epsilon=0}
+{\cal O}\left[T^2\right],
\label{eq:eta-lowT}
\end{equation}
where $\Delta_0$ is the spectroscopic gap, and
$\mu=(1/\Delta_0)|d\Delta(\phi)/d\phi|_{\phi_{\sf node}}$ is the
opening rate of the gap function at the node. In analogy with the
thermal conductivity \cite{LF05}, the remaining effect at $T\ll T^*$
of inelastic scattering is a modification of the $T\rightarrow 0$
asymptotic. When we write explicitly ${\cal N}_f\rightarrow {\cal
N}_f^*={\cal N}_f^0(1+\lambda_{\sf in})$ and $v_f\rightarrow
v_f^*=v_f^0/(1+\lambda_{\sf in})$ in Eq.~(\ref{eq:eta-lowT}), one
factor $1+\lambda_{\sf in}$ remains in the denominator. Within the
bare theory, this result can be traced back to that the spectroscopic
gap $\Delta_0$ in the weak-coupling limit is replaced by the strong
coupling off-diagonal function $W(0)$.

In summary, we have discussed the temperature-dependence of the
thermoelectric effect that results from the interplay of inelastic
electron-boson scattering and elastic impurity scattering. At high
temperatures, $T>T^*$, particle-hole symmetric inelastic scattering
dominates and the thermoelectric coefficient is suppressed. On the
other hand, at low temperatures, $T<T^*$, particle-hole asymmetric
elastic impurity scattering dominates and the thermoelectric
coefficient is enhanced.

%\begin{theacknowledgments}
We gratefully acknowledge financial support from the Swedish Research
Council (M.F), and the EC under the Spintronics Network
RTN2-2001-00440 (T.L).
%\end{theacknowledgments}
\vspace*{-0.4truecm}
\bibliographystyle{aipproc}   % if natbib is available

\end{document}